\documentclass[12pt]{article}
\usepackage{graphicx}
\begin{document}
\centerline{\bf Monte Carlo Simulation of Age-Dependent Host-Parasite Relations}

\bigskip

Dietrich Stauffer$^{1,2}$, Ana Proykova$^1$ and Karl-Heinz Lampe$^3$

\bigskip
\noindent
$^1$ Department of Atomic Physics, University of Sofia, Sofia-1126, Bulgaria.

\noindent
$^2$ Institute for Theoretical Physics, Cologne University, 
D-50923 K\"oln, Euroland.

\noindent
$^3$ Zoologisches Forschungsmuseum Alexander Koenig, Adenauerallee 168,
D-53113 Bonn, Germany.

\bigskip

Abstract: The death of a biological population is an extreme event which we 
investigate here for a host-parasitoid system. Our simulations using the Penna 
ageing model show how biological evolution can ``teach'' the parasitoids to 
avoid extinction by waiting for the right age of the host. We also show the
dependence of extinction time on the population size.

\bigskip

The death of a biological population is an extreme event \cite{jentsch}
which we investigate here for a host-parasitoid system. (Parasites let the host 
survive, parasitoids eventually kill the host. In the following both are called
parasites.) For seasonal hosts (plants or 
animals) the parasites can survive only if they parasitise 
the right developmental stage (egg, larva, pupa, imago) of their host. 
Therefore they have to attack their host at the right time, i.e.  at the right 
age of the host or the respective right month of the year \cite{lampe}. 
Earlier the host has not yet accumulated enough resources, unsuitable for an 
attack of a parasite. Biological evolution thus should ensure that 
the parasites ``know'' what the proper age of the host or time of the year is.
The present note aims to check if such an evolutionary self-organization 
can be simulated; we do not deal with the process how the parasites recognise
the age of the host or the month of the year.

\begin{figure}[hbt]
\begin{center}
\includegraphics[angle=-90,scale=0.5]{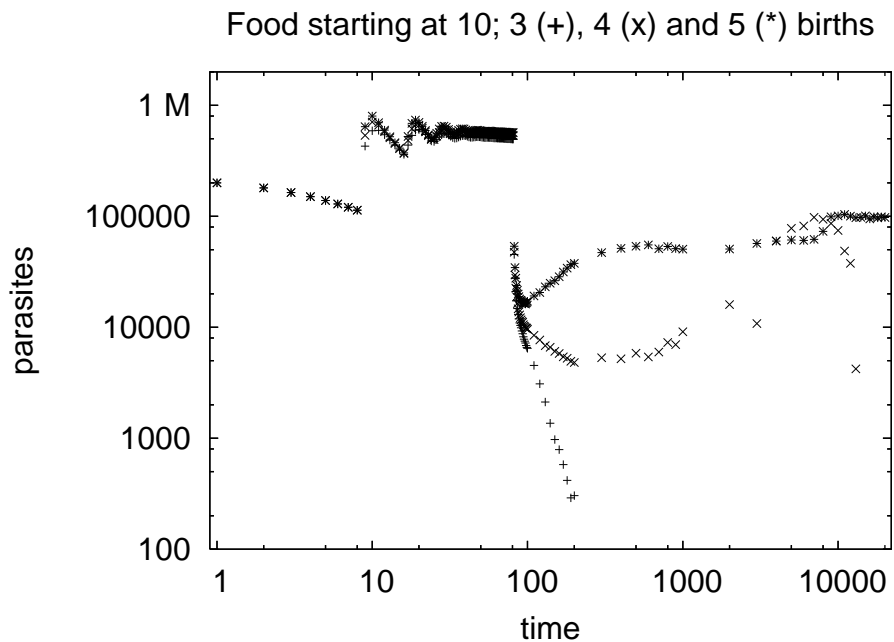}
\end{center}
\caption{Time dependence of number of parasites near threshold for extinction,
for birth rates 3, 4 and 5.
}
\end{figure}
% 48nn, 48n, 45n  u 1:4

\begin{figure}[hbt]
\begin{center}
\includegraphics[angle=-90,scale=0.5]{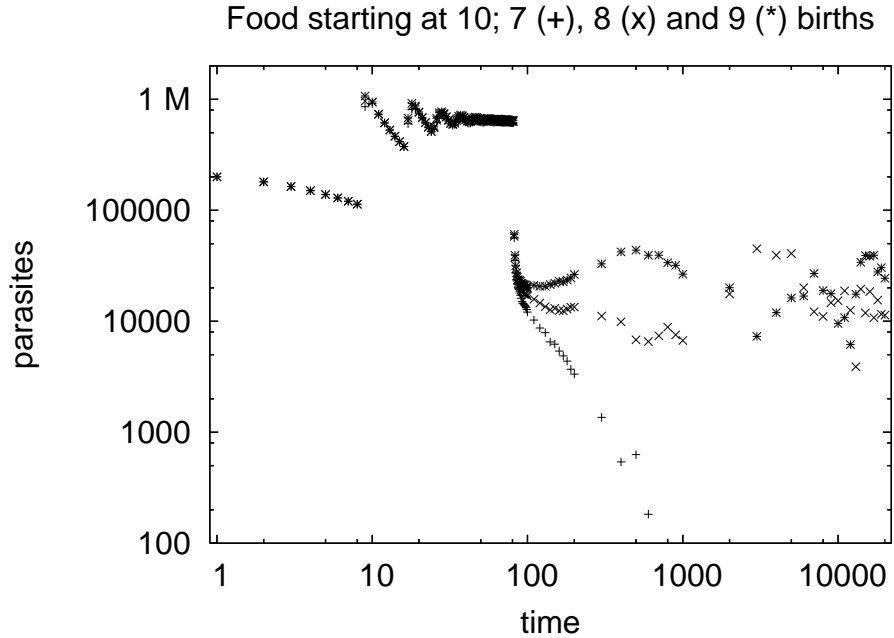}
\end{center}
\caption{As previous figure but with continuous mutations of desired host ages,
and birth rates 7, 8 and 9.
}
\end{figure}
% 43nn, 43n, 44n u 1:4

\begin{figure}[hbt]
\begin{center}
\includegraphics[angle=-90,scale=0.5]{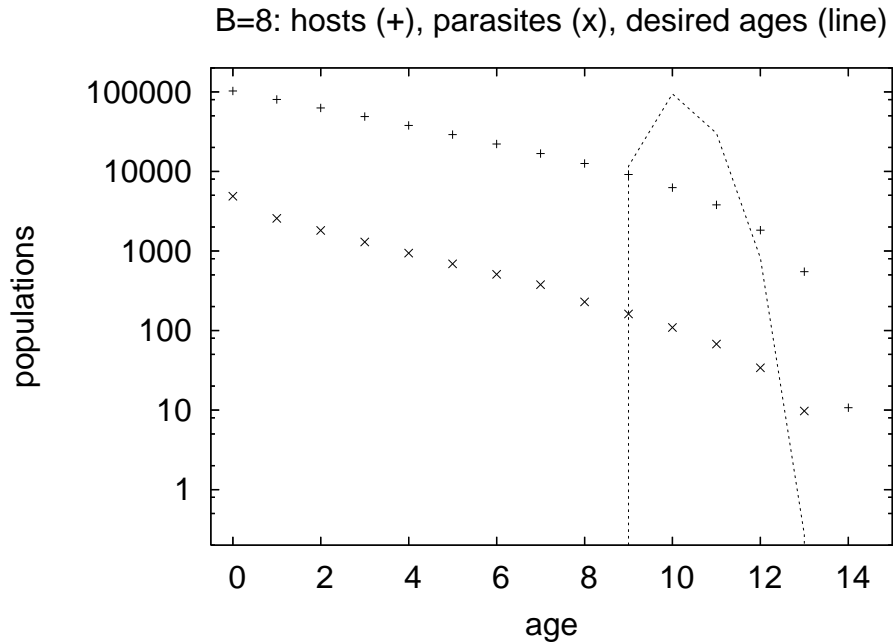}
\end{center}
\caption{Summed age distribution of hosts (+) and parasites(x), and histogram
(line, divided by 1000) of desired age, for continuous mutations of desired food
ages. Without these mutations the histogram is nonzero only at age = 10, while
the age distribution is about the same. Initially, the histogram is flat between
ages 0 and 31.
}
\end{figure}
%43n

\begin{figure}[hbt]
\begin{center}
\includegraphics[angle=-90,scale=0.5]{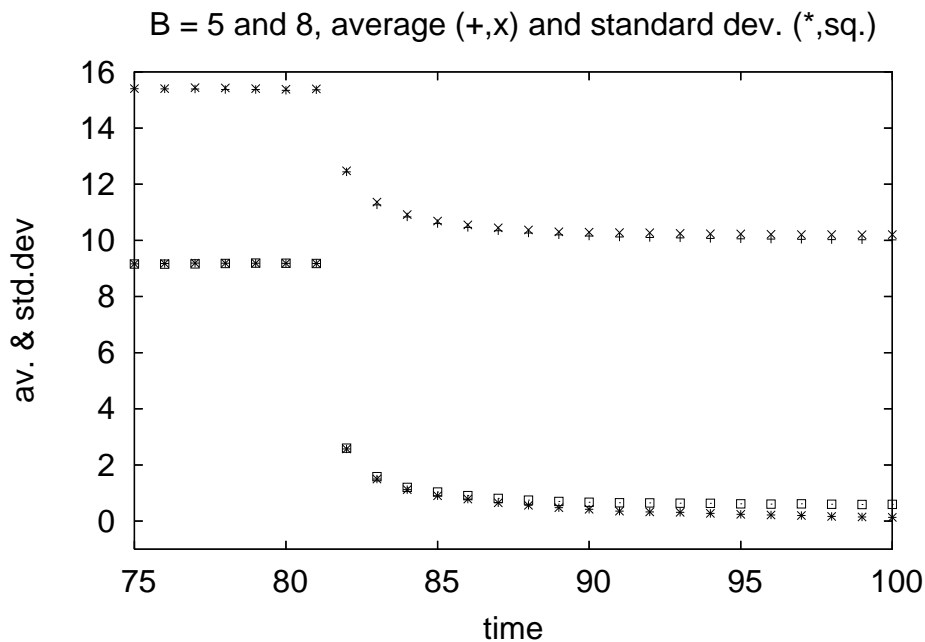}
\end{center}
\caption{Time dependence of the average desired age and its standard deviation
shortly before and after the parasites start to select hosts according to
host age. Birth rate 5 without and birth rate 8 with continuous mutation of
desired host age; these mutations keep the standard deviations above zero.
}
\end{figure}
% 43n 45n

\begin{figure}[hbt]
\begin{center}
\includegraphics[angle=-90,scale=0.55]{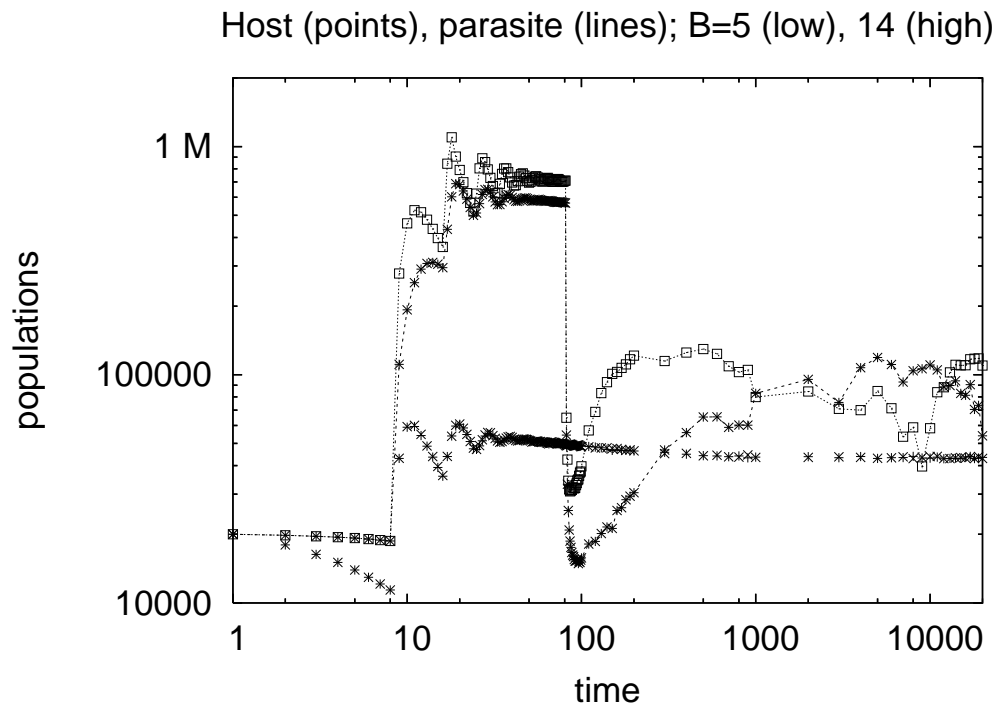}
\end{center}
\caption{Time dependence of the number of hosts (separate symbols) and parasites
(lines with stars and squares) without (B=5,*) and with (B=17,sq.) continuous
mutations of the desired host ages.
}
\end{figure}
% 56 57

\begin{figure}[hbt]
\begin{center}
\includegraphics[angle=-90,scale=0.55]{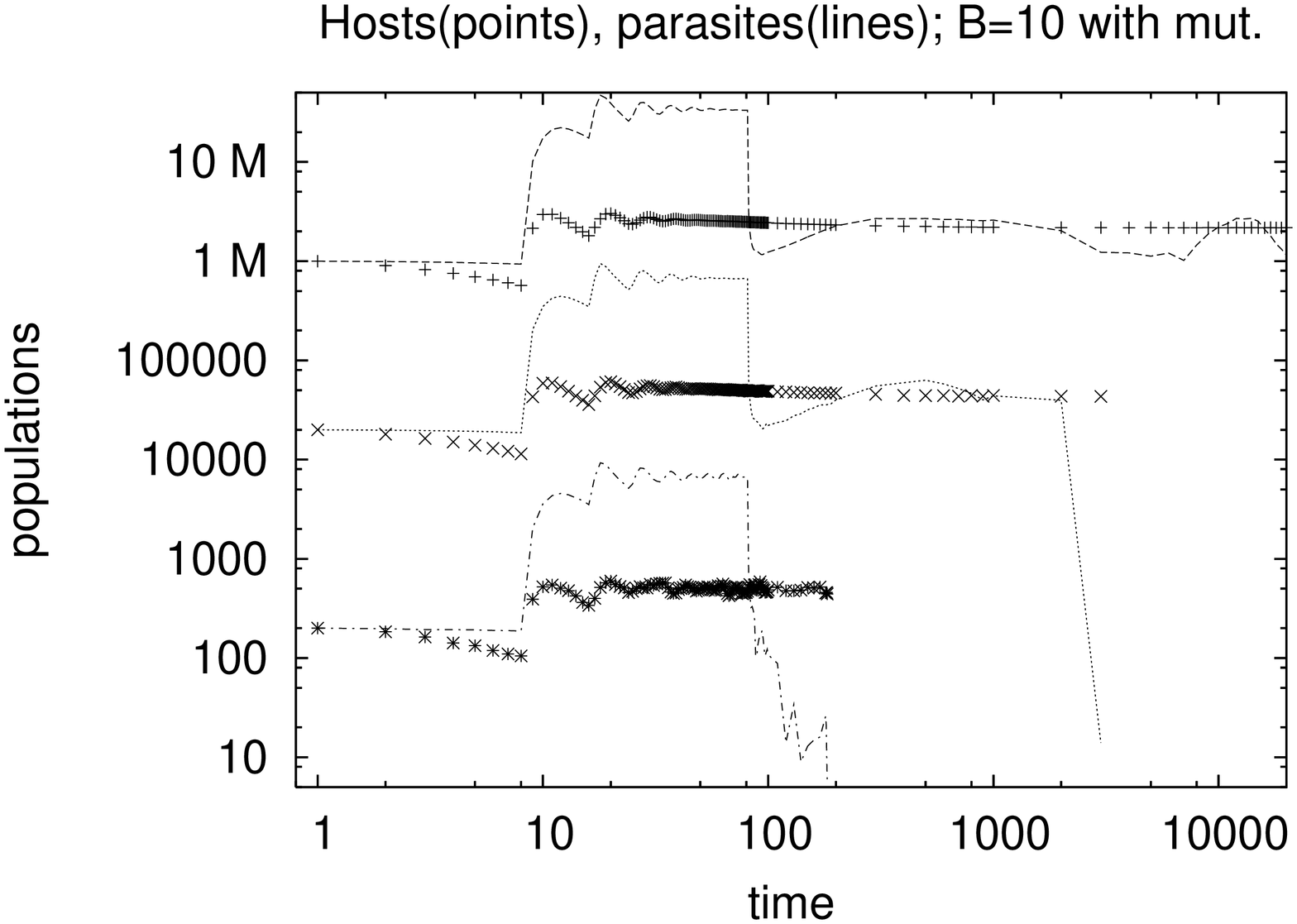}
\end{center}
\caption{Time dependence of the number of hosts (separate symbols) and parasites
(lines with stars and squares) with continuous
mutations of the desired host ages and $B = 10$ close to the extinction value.
Three population sizes are showsn, differeing by two and four decades.
}
\end{figure}
% 58

One example are diversified host-parasite-systems: Endoparasites
deposit their eggs in the interior of the host animal, and thus in case
of young host larvae for example they have to wait 
until the host has developed enough. Then they compete with ectoparasites
attacking the host animal from the outside. Thus the consideration of the
age of the individuals is important, similar to the 'just in time' philosophy
in logistics. We will simulate both fluctuations in equilibrium (normal case),
and possible extinction (extreme event) and thus will find the border between 
extinction and survival. 

We use the standard asexual Penna ageing model of mutation accumulation, with 
reproduction starting an age of 8 time intervals, using a ``genome'' of 32 bits 
representing at most 32 time intervals of the life, with 3 set active bits 
(bad mutations) killing the individual, and a Verhulst factor killing at
all ages because of overpopulation. At each iteration, each mature host
produces three and each mature parasite produces $B$ offspring. At each
birth, one random mutation is made for all offspring by flipping a randomly
selected bit from 0 (healthy) to 1 (sick). (If it is 1 already it stays at 1:
no new mutation.) Typically, $10^4$ iterations were 
averaged over after the populations had reached roughly a dynamic equilibrium 
of births and deaths. More details on the Penna model of 1995 are given in many 
articles and two books \cite{books}, The border between extinction and 
survival was investigated in \cite{malarz} without host-parasite relations.

With this standard model, hosts and parasites are simulated together, using 
separate Verhulst factors with the same carrying capacity. In contrast to the 
hosts, the parasites at each iteration make 100 attempts to invade a host 
of their desired age. (These desired host ages are at the beginning distributed 
randomly
between zero and 31.) If all 100 attempts were unsuccessful, they die. Otherwise
they have enough food provided the host age is at least 10, slightly above
the minimum reproduction age of 8. If the host is too young, then again the 
parasite dies; if not it survives and ages. 

In some simulations we also mutated, randomly and continuously,
the desired host age. Then at birth of a parasite, not only the usual 
bit-string genome is mutated but also its desired host age. With 25 percent 
probability it increases by one unit, with 25 percent probability it decreases
by one unit, and in the remaining 50 percent of the cases it stays constant.

Parasites do not necessarily affect the health of the host \cite{lampe}, and we 
neglect this influence completely. Generally, if we look at the boundary between
survival and extinction of the parasites, the influence of the parasites
becomes zero at this boundary.
 
In the simulations we first make 80 iterations without coupling hosts and 
parasites, in order to get close to a stationary state; see left parts
of Figs.1,2. From then on the 
parasites need a host of the proper age, and thus most of them die. The 
question is whether some survive nevertheless and replenish the parasite 
population.

Figure 1 shows the time dependence of the populations without and Fig.2 with
the continuous mutations of the desired host ages. With these mutations, Fig.3
shows the age distributions (as usual for the Penna model) and the distribution
of the desired host ages. The latter one changes from its initial flat shape
between the possible ages zero and 31, to a narrow peak close to the minimum
age of 10 where hosts become useful for the parasites. If the mutations of the
desired host ages are omitted, the distribution becomes a delta function at
age = 10. On an expanded time scale, Fig.4 shows the rapid transition from 
a flat to a sharply peaked distribution, within a few iterations.
Usually in reality, as opposed to the above simulations, the parasites are 
smaller and more numerous than the hosts. We now achieve this by making the
parasite carrying capacity 10 times bigger than that for the hosts, Fig.5.

When we demand that the host age is not only at least 10 but exactly a fixed
age than this age has to be fixed at unrealistically low ages (not shown).
 
Small populations subject to random fluctuations always die out if observed long
enough\cite{pal}. ``Small'' means here than the fluctuations in the population
size are much smaller than the average population size. Fig.6 shows in a 
comparison of three population sizes that the host populations fluctuate
relatively less when the population size is increased, and do not die out. 
The parasite populations, on the other hand, behave differently, and die out 
even when on average they number tens of thousands. The reason is that in 
contrast to Fig.5 with $B=17$ we took for Fig.6 a lower birthrate of 10, close
to the boundary between survival and extinction of the parasites.

In summary, we simulated how natural evolution via survival of the fittest 
can ``teach'' the parasites to avoid extinction by attacking hosts at the
proper host age (or time of the year for seasonal hosts). In the future one 
could simulate more explicitely different life strategies of the parasites, 
e.g. endoparasitoids versus true parasite for plants or animals.

This note is dedicated to Naeem Jan with whom DS started ageing simulations.

\end{document}